\documentclass[preprint, authoryear]{elsarticle}
\usepackage[T1]{fontenc}
\usepackage[utf8]{inputenc}
\usepackage{geometry}
\usepackage{amsmath}
\usepackage{amsthm}
\usepackage{graphicx}
\usepackage{listings}

\makeatletter
\def\ps@pprintTitle{%
 \let\@oddhead\@empty
 \let\@evenhead\@empty
 \def\@oddfoot{}%
 \let\@evenfoot\@oddfoot}
\makeatother

\theoremstyle{plain}
\newtheorem{thm}{\protect\theoremname}
\theoremstyle{definition}
\newtheorem{example}[thm]{\protect\examplename}
\providecommand{\examplename}{Example}
\providecommand{\theoremname}{Theorem}

\begin{document}

\begin{frontmatter}
\title{Correcting for attenuation due to measurement error}
\author{Jonas Moss}
\ead{jonasmgj@math.uio.no}
\address{Department of Mathematics, University of Oslo, PB 1053, Blindern, NO-0316, Oslo, Norway}
\begin{abstract}
I present a frequentist method for quantifying uncertainty when correcting
correlations for attenuation due to measurement error. The method
is conservative but has far better coverage properties than the methods
currently used when sample sizes are small. I recommend the use of
confidence curves in favor of confidence intervals when this method
is used. I introduce the R package ``attenuation'' which can be
used to calculate and visualize the methods described in this paper.
\end{abstract}
\end{frontmatter}

\section{Introduction}

Here is a story about two researchers named Alice and Bob. Alice wants
to calculate the correlation $\rho$ between $X,Y$, but all she has
is the correlation $\rho_{x'y'}$ between two noisy measurements $X'$
and $Y'$. How can she recover the correlation between $X$ and $Y$?
If Alice knows the squared correlations, also known as the reliabilities, $r_{xx'}^{2}$ and $r_{yy'}^{2}$
for the measurements $X'$ and $Y'$, she can use the classical formula
of \citet{spearman1904proof}, $\rho=\rho_{y'x'}/r_{xx'}r_{yy'}$.
Bob is in a worse position, as he does not know $\rho_{x'y'}$, $r_{xx'}^{2}$
and $r_{yy'}^{2}$, he only has estimates of them. His estimate of
$\rho_{x'y'}$ is $r_{x'y'}$, the sample correlation calculated from
a sample of $n_{1}$ participants. His estimate of $r_{xx'}^{2}$
is $\widehat{\alpha_{xx'}}$, the maximum likelihood estimator of
coefficient alpha \citep[equation 1]{mcneish2018thanks} calculated
from a sample of $n_{2}$ participants and $k_{2}$ testlets. His
estimate of $R_{yy'}$ is $\widehat{\omega_{yy'}}$, McDonald's omega
\citep[equation 2]{mcneish2018thanks} based on $n_{3}$ participants
and $k_{3}$ testlets. His uses the plug-in estimator of $\rho_{x'y'}$
using Spearman's formula, and obtains $r_{x'y'}/\widehat{\alpha_{xx'}}^{1/2}\widehat{\omega_{yy'}}^{1/2}>1$.

My purpose with this paper is to help Bob -- and people like him
-- in quantifying the uncertainty of their plug-in estimates of $\rho$.
I propose confidence sets, \emph{p}-values, and confidence curves
that take the variability in all three estimates into account. This
is particularly important when the magnitude of estimate of $\rho$
exceeds $1$, as this can lead to unwarranted conclusions such as
$X$ and $Y$ measure essentially the same thing when it is equally
well explained by sampling variability.

It is hard to construct\emph{ p}-values for $\rho$ since we do not
directly observe the data from the bivariate distribution $\left(X,Y\right)$.
This makes standard sampling theory difficult to use \citep{hakstian1988inferential}.
However, as it turns out it is, it is not too hard to construct approximate
\emph{p}-values if we allow them to be somewhat conservative and do
not attempt to identify the sampling distributions of $\widehat{\rho}$.

The most important confidence set for corrected correlations in the
Hunter-Schmidt confidence interval, see \citet[p. 96 - 103]{hunter2004methods}.
This confidence set is based on normal sampling theory for the correlation
coefficient $r_{x'y'}$ and an assumption that $r_{xx'}^{2}$ and
$r_{yy'}^{2}$ are known. It equals
\begin{equation}
\frac{r_{x'y'}}{r_{xx'}r_{yy'}}\pm\frac{1-r_{x'y'}^{2}}{\left(N-1\right)^{1/2}r_{xx'}r_{yy'}}\label{eq:HS interval}
\end{equation}
where $N$ is the sample size of $r_{x'y'}$. The corresponding \emph{p}-value
for $H_{0}:\rho=\rho^{0}$ is 
\[
2\Phi\left(-\left|r_{x'y'}-\rho^{0}r_{xx'}r_{yy'}\right|,\left(1-r_{x'y'}^{2}\right)\left(N-1\right)^{1/2}\right)
\]
In addition, \citet{padilla2012correlation} proposed a bootstrap
confidence interval, but this is also based on the assumption that
the reliabilities $r_{xx'}^{2}$ and $r_{yy'}^{2}$ are known. \citet{charles2005correction}
has a lengthy discussion of confidence sets and includes some new
constructions as well.

In section \ref{sec:The Method} I how to construct \emph{p}-values,
confidence sets and confidence curves for $\rho$. In the following
section \ref{sec:Confidence sets} I run some simulations investigating
the coverage of the confidence sets. Section \ref{sec:attenuation}
is devoted to a description of $\mathtt{attenuation}$, an R package
devoted to the calculation and visualization of the methods described
in his paper. I briefly conclude in section \ref{sec:Concluding Remarks}.

\section{The Method\label{sec:The Method}}

The underlying model is
\begin{eqnarray}
\left(X,Y\right) & \sim & N\left(0,\left[\begin{array}{cc}
1 & \rho\\
\rho & 1
\end{array}\right]\right)\label{eq:Model}\\
X' & \sim & N\left(X,\sigma_{x}\right)\nonumber \\
Y' & \sim & N\left(Y,\sigma_{y}\right)\nonumber 
\end{eqnarray}
Here $\left(X,Y\right)$ are standardized to make the model identifiable.
The standard deviations $\sigma_{x},\sigma_{y}$ are noise levels
of the measurements $X'$ and $Y'$. The model for $X'$ and $Y'$
are taken from true score theory. The reliability of $X'$ is defined
as $\textrm{Var}\left(X\right)/\textrm{Var}\left(X'\right)=$$\left(1+\sigma_{x}\right)^{-1}=\rho_{xx'}^{2}$,
and likewise for $Y'$. Since $\textrm{Cor}\left(X',X\right)=\left(1+\sigma_{x}\right)^{-1/2}>0$,
the correlation between $X$ and $X'$ is positive, and the correlation
between $Y$ and $Y'$ is positive too.

I will denote $\rho_{x'y'}=\rho_{1}$ , $\rho_{xx'}=\rho_{2}$ and
$\rho_{yy'}=\rho_{3}$ for readability. The estimate $\widehat{\rho_{1}}=r_{1}$
is a sample correlation based on $N_{1}$ observations from a bivariate
normal with true correlation $\rho_{1}$. For now, $\widehat{\rho_{2}}=r_{2}$
and $\widehat{\rho_{3}}=r_{3}$ are sample correlations from bivariate
normals with sample sizes $N_{2}$ and $N_{3}$ and true correlations
$\rho_{2}$ and $\rho_{3}$. I will let $\widehat{\rho_{2}}$ and
$\widehat{\rho_{3}}$ be alpha coefficients later on, in subsection
\ref{subsec:Cronbach alpha p-value}. I will make use of the shorthands
$r=\left(r_{1},r_{2},r_{3}\right)$ and $N=\left(N_{1},N_{2},N_{3}\right)$.

Consider the following testing problem.

\begin{eqnarray}
H_{0} & : & \rho=\rho^{0}\label{eq:Main hypothesis test}\\
H_{1} & : & \rho\neq\rho^{0}\nonumber 
\end{eqnarray}
Notice that the null hypothesis is composite, as $\rho=\rho^{0}$
if and only if our observations $r,N$ are sampled from the probability
$P_{\overline{\rho}}$, where $\overline{\rho}=\left(\rho_{1},\rho_{2},\rho_{3}\right)$
and $\rho^{0}=\rho_{1}/\left(\rho_{2}\rho_{3}\right)$. An hypothesis
test of level $\alpha$ for this problem is defined by an acceptance
set $A_{\rho^{0}}$ satisfying $P_{\overline{\rho}}\left(\omega\in A_{\rho^{0}}\right)\geq1-\alpha$
whenever $\rho^{0}=\rho_{1}/\left(\rho_{2}\rho_{3}\right)$, where
$\overline{\rho}=\left(\rho_{1},\rho_{2},\rho_{3}\right)$.

In order to construct such a set, I will start out with creating a
reasonable size $\alpha$ acceptance set for the simple null hypothesis
\begin{eqnarray}
H_{0} & : & \rho_{1}=\rho_{1}^{0},\rho_{2}=\rho_{2}^{0},\rho_{3}=\rho_{3}^{0}\label{eq:Simple hypothesis}
\end{eqnarray}

To do this, use the \citet{fisher1915frequency} transform to approximate
$$s=\left(\textrm{artanh}\left(r_{1}\right),\textrm{artanh}\left(r_{2}\right),\textrm{artanh}\left(r_{3}\right)\right)$$
as a multivariate normal with mean $$\eta=\left(\textrm{artanh}\left(\rho_{1}\right),\textrm{artanh}\left(\rho_{2}\right),\textrm{artanh}\left(\rho_{3}\right)\right)$$
and diagonal covariance matrix $D$ with diagonal elements $D_{11}=\left(N_{1}-3\right)^{-1}$,
$D_{22}=\left(N_{2}-3\right)^{-1}$, and $D_{33}=\left(N_{3}-3\right)^{-1}$.
Then the smallest set of probability $1-\alpha$ is the interior of
a level set ellipsoid, which equals $A_{\overline{\rho}}=\left\{ \left(\eta-s\right)^{T}D^{-1}\left(\eta-s\right)\leq\chi_{3,1-\alpha}^{2}\right\} $
where $\chi_{3,1-\alpha}^{2}$ is the $1-\alpha$ quantile of a $\chi^{2}$
with three degrees of freedom. To test the hypothesis (\ref{eq:Simple hypothesis}),
just check if $r$ is included in $A_{\overline{\rho}}$.

Now let us return to our the testing problem \ref{eq:Main hypothesis test}.
Define the acceptance set by $A_{\rho^{0}}=\bigcup A_{\overline{\rho}}$,
where the union is over all $\overline{\rho}=\left(\rho_{1},\rho_{2},\rho_{3}\right)$
such that $\rho^{0}=\rho_{1}/\left(\rho_{2}\rho_{3}\right)$. Then
$P_{\overline{\rho}}\left(\omega\in A_{\rho^{0}}\right)\geq P_{\overline{\rho}}\left(A_{\overline{\rho}}\right)\geq1-\alpha$,
hence it is an acceptance set of a level $\alpha$ test of (\ref{eq:Main hypothesis test}).

Since the ellipses $A_{\overline{\rho}}$ are nested as a function
of $\alpha$, the acceptance sets $A_{\rho^{0}}$ are nested as a
function of $\alpha$ too. This implies there is a \emph{p}-value
with $\left\{ A_{\rho^{0}}\right\} $ as underlying acceptance sets,
see e.g. \citet[p. 63]{lehmann2006testing}.

The \emph{p}-value at $\rho$ is the solution to the following program

\begin{align}
\textrm{maximize} &  & 1-F_{\chi_{3}^{2}}\left[\left(\eta-s\right)^{T}D^{-1}\left(\eta-s\right)\right]\label{eq:p-value program}\\
\textrm{subject to} &  & \rho_{1}/\left(\rho_{2}\rho_{3}\right)=\rho\nonumber \\
 &  & \rho_{1}\in\left[-1,1\right],\rho_{i}\in\left[0,1\right],i=2,3\nonumber 
\end{align}
The inequality constraints $\rho_{i}\in\left[0,1\right],i=2,3$ are
imposed to make the \emph{p}-value consistent with the model \ref{eq:Model}.

Since $\rho_{1}/\left(\rho_{2}\rho_{3}\right)=\rho$ if and only if
$\rho_{1}=\rho\rho_{2}\rho_{3}$, this is can be rewritten as maximization
problem of two parameters constrained to the unit interval. Moreover,
since $F_{\chi_{3}^{2}}$ is strictly increasing, the maximizer of
the program \ref{eq:p-value program} is the same the minimizer

\begin{align}
\textrm{minimize} &  & \left(\eta-s\right)^{T}D^{-1}\left(\eta-s\right)\nonumber \\
\textrm{subject to} &  & \rho_{1}\in\left[-1,1\right],\rho_{i}\in\left[0,1\right],i=2,3\label{eq:p-value program, minizer}\\
 &  & \eta_{1}=\text{\textrm{artanh}}\left(\rho\rho_{2}\rho_{3}\right)\nonumber 
\end{align}
As this is a strictly convex program it has a unique solution $\left(\rho'_{2},\rho'_{3}\right)$.
It is easy to solve with numerical optimization procedures such as
the $\mathtt{optim}$ function of $\mathtt{R}$ \citep{Rlang}. To
recover the \emph{p}-value, simply plug the solution $\eta=\left(\text{\textrm{artanh}}\left(\rho\rho'_{2}\rho'_{3}\right),\text{\textrm{artanh}}\left(\rho'_{2}\right),\text{\textrm{artanh}}\left(\rho'_{3}\right)\right)$
into $1-F_{\chi_{3}^{2}}\left[\left(\eta-s\right)^{T}D^{-1}\left(\eta-s\right)\right]$.

\subsection{Confidence Curves}

\begin{figure}
\noindent \begin{centering}
\includegraphics[scale=0.35]{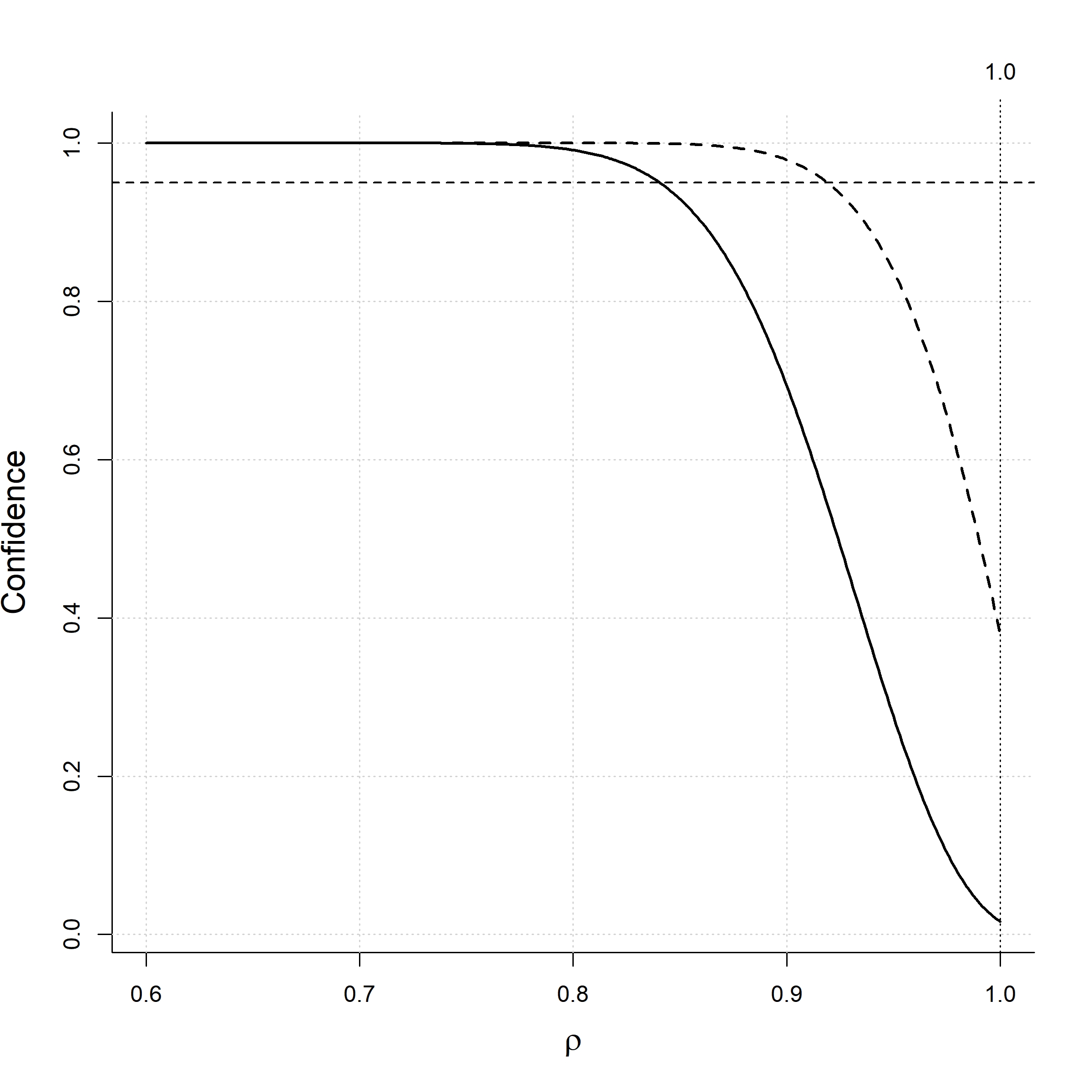}\includegraphics[scale=0.35]{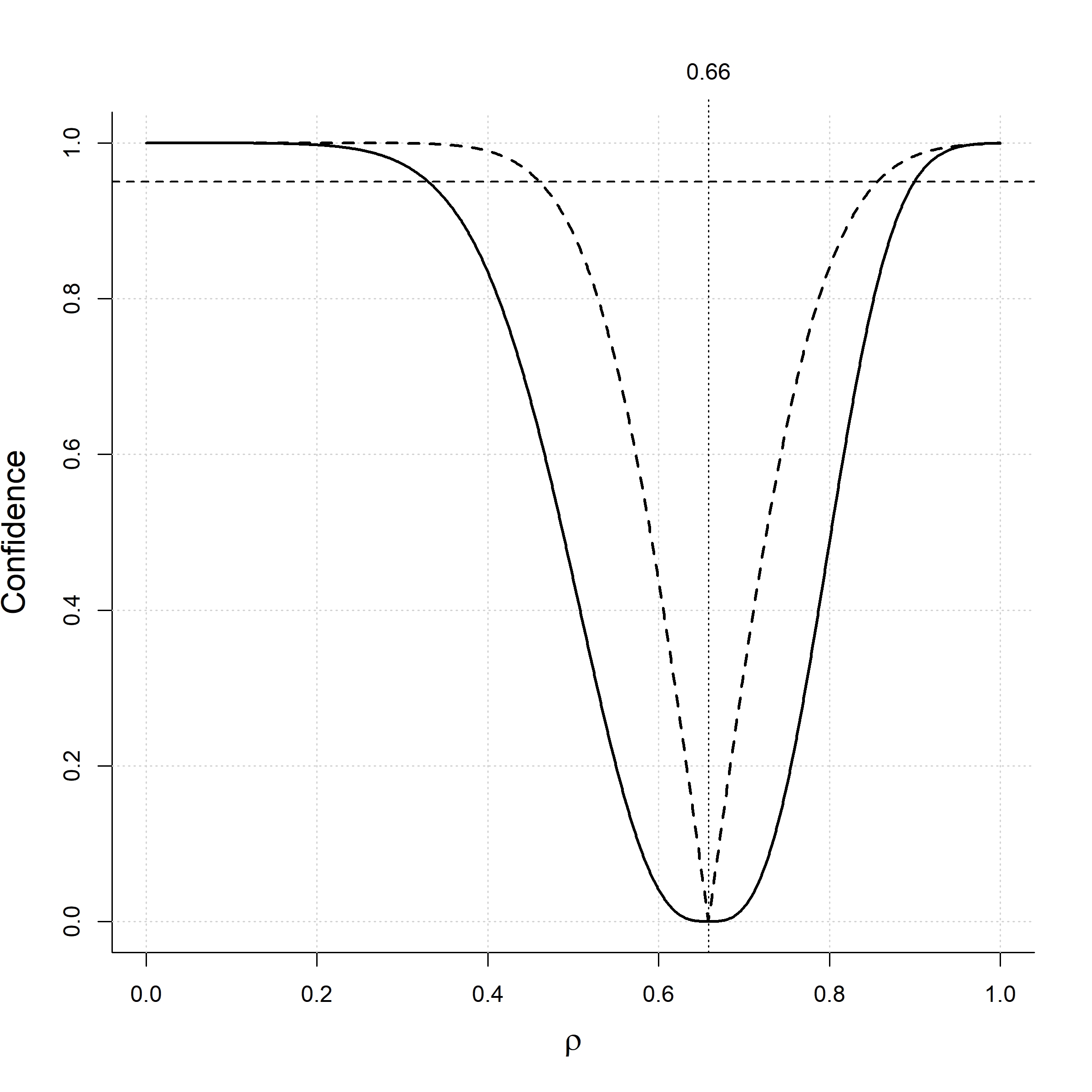}
\par\end{centering}
\caption{\label{fig:Confidence curves} Confidence curves with the $95\%$
confidence set and point estimates emphasized. The solid curves are
calculated by the new method and the dashed curve by the Hunter-Schmidt
method. . \textbf{(left)} Corrected correlation between Gordon and
Piers-Harris from \citet{marx1978construct}. \textbf{(right) }Corrected
correlation between Facilitation and Agreeableness from \citet{fiori2012selective}.}
\end{figure}

A confidence curve \citep{birnbaum1961confidence,schweder2016confidence},
also known as a \emph{p}-value function \citep{martin2017statistical},
is a the function $\rho\mapsto1-p_{\rho}\left(r,N\right)$, where
$p_{\rho}\left(r,N\right)$ is the \emph{p}-value at $\rho$ calculated
under the data $\left(r,N\right)$. From a confidence curve you can
read all $1-\alpha$ level confidence sets and a point estimate as
the minimizer of the curve. Confidence curves are particularly useful
for understanding the uncertainty in $\rho$ since the confidence
sets can potentially be either empty or cover the entire interval
$\left[-1,1\right]$. If you come across a level $\alpha$ confidence
set that is empty, you would probably try to calculate a confidence
set with a lower level, say $\alpha/2$, and check if it is non-empty.
But such a procedure is unprincipled. By using confidence curves,
you do not need to make a choice of $\alpha$ for yourself and your
readers.
\begin{example}
\label{exa:CC examples} \citet{marx1978construct} studied three
self-report measures of self-concept on $488$ six-graders. In the
results section they provide sample correlations between the three
measures and their reliabilities as sample Cronbach alphas. The correlation
between the measure of self-concept called \emph{Gordon} and the measure
of self-concept called\emph{ Piers-Harris} is $r_{1}=.57$ with reliabilities
$r_{2}^{2}=.56$ and $r_{3}^{2}=.55$. Using Spearman's formula yields
an estimate of $\rho$ equal to $1.03$, which is impossible. The
confidence curve for this data is to the left in figure, where the
solid curve is the new method and the dashed curve is the Hunter-Schmidt
method (\ref{eq:HS interval}). \ref{fig:Confidence curves}. The
$95\%$ confidence set is $\left[0.84,1\right]$ for the new method
and $\left[0.92,1\right]$ for the Hunter-Schmidt method.
\end{example}

There is no need for the correlations to come from the same study
or have the same sample sizes.
\begin{example}
\citet[table 1]{fiori2012selective} contains sample correlations
$(n=85)$ between the branches of the MSCEIT test of emotional intelligence
\citep{mayer2002mayer} and the dimensions of The Big Five Inventory
(BFI) \citep{john1999big}. The sample correlation between the Facilitating
branch of the MSCEIT and the Agreeableness dimension of the BFI is
$.52$. \citet{mayer2003measuring} provides an estimate of coefficient
alpha $(.79)$ for Facilitating with $n=2028$, while \citet{benet1998cinco}
has an estimate of Cronbach alpha for Agreeableness equal to $0.79$
with $n=711$. The confidence curve for this data is to the right
in figure, where the solid curve is the new method and the dashed
curve is the Hunter-Schmidt method. \ref{fig:Confidence curves}.
The $95\%$ confidence set is $\left[0.33,0.90\right]$ for the new
method and $\left[0.46,0.86\right]$ for the Hunter-Schmidt method.
\end{example}

\subsection{Using Cronbach's $\alpha$\label{subsec:Cronbach alpha p-value}}

In the previous section I assumed that $r_{2}$ and $r_{3}$ were
sample correlations. But such correlations are hard to come by, since
the latent $X,Y$ are almost always unknown. Instead, the reliabilities
are estimated indirectly using typically coefficient alpha, which
is by far most popular measure of reliability in the psychological
literature \citep{mcneish2018thanks}. Coefficient alpha does not
have the same sampling distribution as $r$, so we cannot expect the
\emph{p}-values to be equals. Luckily, it is easy to modify the \emph{p}-value
program \ref{eq:p-value program} to work for coefficient alpha.

The essential ingredient is the formula for the asymptotic distribution
of coefficient alpha by \citet{van2000distribution}:

\begin{equation}
n^{1/2}\left[\frac{1}{2}\log\left(1-r^{2}\right)-\frac{1}{2}\log\left(1-\alpha\right)\right]\stackrel{d}{\to}N\left(0,\frac{k}{2\left(k-1\right)}\right)\label{eq:Cronbach's alpha, asymptotic distribution}
\end{equation}
were $\alpha$ is coefficient alpha, $r^{2}$ is its maximum likelihood
estimator of $\widehat{\alpha}$, $n$ is the sample size and $k$
is the number of testlets. This result holds under the assumption
of multivariate normality and compound symmetry of the covariance
matrix of the testlets. Another name for the compound symmetry assumption
is that tests are parallell.

The modification of program \ref{eq:p-value program, minizer} reads

\begin{align}
\textrm{minimize} &  & \left(\eta-s\right)^{T}D^{-1}\left(\eta-s\right)\nonumber \\
\textrm{subject to} &  & \rho_{1}\in\left[-1,1\right],R_{i}\in\left[0,1\right],i=2,3\label{eq:p-value program, Cronbach's alpha, minizer}\\
 &  & \eta_{1}=\text{\textrm{artanh}}\left(\rho r_{2}r_{3}\right)\nonumber 
\end{align}
where $\eta_{i}=1/2\log\left(1-r_{i}^{2}\right)$ for $i=2,3$ and
$D$ is a diagonal matrix with elements $D_{1}=\left(N_{1}-3\right)^{1/2}$,
$D_{2}=\left[2N_{2}\left(k_{2}-1\right)/k_{2}\right]^{1/2}$ and $D_{3}=\left[2N_{3}\left(k_{3}-1\right)/k_{3}\right]^{1/2}$,
and $r_{2},r_{3}$ are the positive roots of $r_{2}^{2}$ and $r_{3}^{2}$.

\section{Coverage of the Confidence Sets\label{sec:Confidence sets}}

\begin{figure}
\noindent \begin{centering}
\includegraphics[scale=0.35]{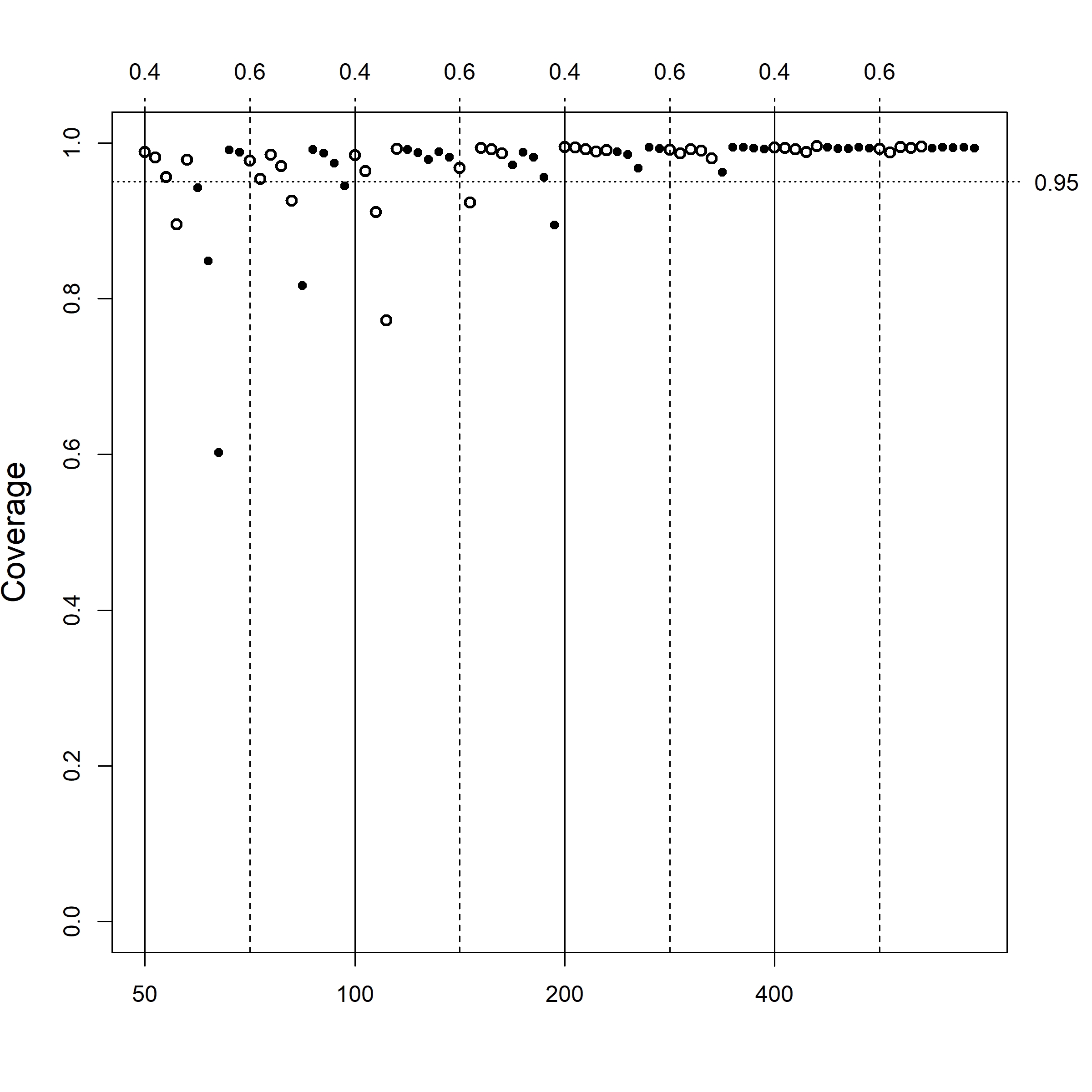}\includegraphics[scale=0.35]{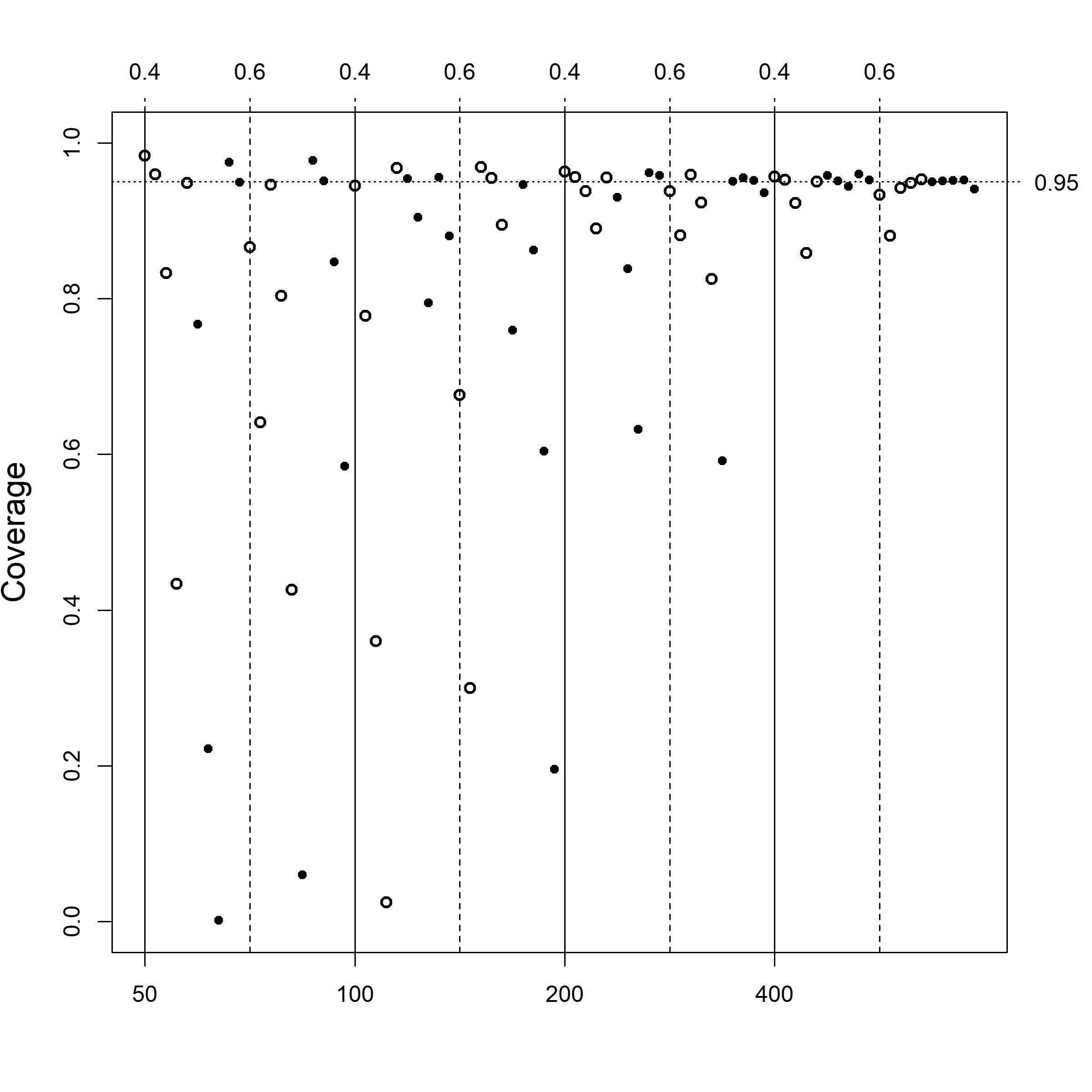}
\par\end{centering}
\caption{\label{fig:Simulations} Simulated coverage for the \textbf{(left)}
confidence sets based on coefficient alpha (mean: $0.970,$ sd: $0.06$)
and \textbf{(right)} the Hunter-Schmidt method (mean: $0.82,$ sd:
$0.23$) . The solid vertical lines delineate the different values
of $n$ and the dashed vertical lines the different values of $\rho$.
The circles correspond to $k=4$, while the solid dots correspond
to $k=8$. The rest of the points are ordered according to $r^{2}\in\left\{ 0.25,0.36,0.49,0.64,0.81\right\} $.
The limits on the $y$-axes are not the same in both plots.}

\end{figure}

In this section I simulate the coverage of the confidence sets based
on correlations (program \ref{eq:p-value program}), coefficient alpha
(program \ref{eq:p-value program, Cronbach's alpha, minizer}) and
the Hunter-Schmidt method \ref{eq:HS interval} using the same setup
as \citet{fan2003two}. This simulation involves four sample sizes
$N$, two different true correlations $\rho$, two different number
of testlets $k$, and five different true reliabilities $r^{2}$:

\begin{eqnarray*}
N & \in & \left\{ 50,100,200,400\right\} \\
\rho & \in & \left\{ 0.4,0.6\right\} \\
k & \in & \left\{ 4,8\right\} \\
R & \in & \left\{ 0.25,0.36,0.49,0.64,0.81\right\} 
\end{eqnarray*}
For each combination of $N,\rho,k,R$ I simulate two coefficient alphas
$r_{2}^{2},r_{3}^{2}$ based on $k$ testlets, a sample size of $N$,
and a true coefficeint alpha of $R$, and one correlation $r=\rho R$
based on a sample size of $N$. I check if the resulting $r,N,k$
are included in the acceptance sets of the three tests at level $\alpha=0.05$.
I repeat each simulation $10000$ times. The results can be found
in the OSF repository of this paper at https://osf.io/54zea/. I do
not show the results for the confidence sets based on correlations
since they are not interesting enough. There is no discernible pattern
in the coverage, with mean $0.99$ and standard deviation $0.0007$.
Figure \ref{fig:Simulations} shows the results of the simulation
for the confidence sets based on coefficient alpha and the Hunter-Schmidt
method.

The coverage of both the confidence sets based on correlations and
the confidence sets based on coefficient alpha are uniformly much
larger than the nominal $0.95$.

The confidence sets based on correlations are conservative for all
sample sizes, which should be no surprise given their construction.
What is more surprising is the poor coverage of the coefficient alpha
confidence set for some parameters when the sample size is low. This
is probably due to slow convergence of the sample coefficient alpha
to the limiting distribution in \ref{eq:Cronbach's alpha, asymptotic distribution}.

The coverage of the confidence sets based on correlations agree well
with the confidence sets based on coefficient alpha when the sample
size is large. Since the confidence sets based on correlations are
easier to calculate, require less information and have better coverage
for small sample sizes, it is reasonable to prefer the confidence
sets based on correlations. Even if some choices of $N,\rho,k,r^{2}$
turn out to make the coverage of the correlation based confidence
set smaller, the conservatism of the confidence set based on correlations
is so large the true coverage of the confidence set is likely to be
larger than the nominal coverage anyway.

The coverage of the Hunter-Schmidt method is bad for sample sizes
smaller than $400$ and horrible when below $200$. On the other hand,
its coverage is good for $n=400$. While it fails to achieve a coverage
close $0.95$ for all parameter values, it is not nearly as conservative
as the new confidence sets.

\section{The $\texttt{attenuation}$ package\label{sec:attenuation}}

The R package $\texttt{attenuation}$ has three core functions, $\texttt{p\_value}$
for calculating \emph{p}-values, $\texttt{cc}$ for calculating confidence
curves, and $\texttt{ci}$ for calculating confidence sets. Each of
these functions support four methods:
\begin{itemize}
\item $\mathtt{corr}$: The method based on sample correlation described
in program \ref{eq:p-value program}.
\item $\mathtt{free}$: The method in \ref{eq:p-value program}, except
that the correlations $\rho_{2},\rho_{3}$ are allowed to be negative.
\item $\mathtt{cronbach}$: The method based on the asymptotic distribution
for coefficient alpha in \ref{eq:p-value program, Cronbach's alpha, minizer}.
\item $\mathtt{HS}$: The Hunter-Schmidt method (\ref{eq:HS interval}).
\end{itemize}
The simulations and examples in this paper were done using the $\texttt{attenuation}$
package. It is available on $\mathtt{CRAN}$. Here is an example calculation
of a confidence set.

\begin{lstlisting}[language=R] 
library("attenuation")
r = c(0.20, sqrt(0.45), sqrt(0.55))
N = c(100, 100, 100)
ci(r, N, method = "corr")
#> [1] -0.1647174 0.9958587
\end{lstlisting}

\section{Concluding Remarks\label{sec:Concluding Remarks}}

My proposed \emph{p}-value \ref{eq:p-value program} is not likely
to be optimal in any sense of the word. Still, it is the result of
a reasonable and intuitive construction, and is the first \emph{p}-value
with good behavior under small sample sizes. I note that I have \emph{not
}proven that the \emph{p}-value has the correct level, as this would
require something along the lines of a proof of uniform convergence
in distribution (in in $\rho$) of $\left(n-3\right)^{1/2}\left(\textrm{artanh}\left(\rho\right)-\textrm{artanh}\left(r\right)\right)$
to $N\left(0,1\right)$.

The\emph{ }method is conservative, giving confidence sets with true
coverage far above the nominal coverage in a simulation that violates
its assumption. It would be nice to have smaller confidence sets,
perhaps by a modification of the method in this paper. It is well
known that the assumptions underlying coefficient alpha as a measure
of reliability (i.e tau equivalence) seldom holds \citep{novick1967coefficient}.
For instance ''A simulation by Green and Yang (2009a) found that
coefficient alpha may underestimate the true reliability by as much
as 20\% when tau equivalence is violated (e.g., if the true reliability
is 0.70, coefficient alpha would estimate reliability in the mid 0.50s).''
\citep[p. 4]{mcneish2018thanks} Since the estimates of the reliability
coefficients are likely to be inconsistent, there is a strong extra-statistical
case in favor of conservatism.

\bibliographystyle{elsarticle-harv}
\bibliography{main.bib}

\end{document}